\documentstyle[aps,epsf,twocolumn,amssymb,graphicx]{revtex}
\begin{document}
\draft
\title{Inter- and Intragranular Effects in Superconducting Compacted Platinum Powders}
\author{Alexander Schindler, Reinhard K\"onig, Thomas Herrmannsd\"orfer and Hans F. Braun}
\address{Physikalisches Institut, Universit\"at Bayreuth, D-95440 Bayreuth, 
Germany}
\date{\today}

\maketitle

\begin{abstract}
Compacted platinum powders exhibit a sharp onset of diamagnetic screening at $T \simeq 1.9$ mK in zero magnetic field in all samples investigated. This sharp onset is interpreted in terms of the intragranular transition into the superconducting state. At lower temperatures, the magnetic ac susceptibility strongly depends on the ac field amplitude and reflects the small intergranular critical current density $j_{c}$. This critical current density shows a strong dependence on the packing fraction $f$ of the granular samples. Surprisingly, $j_{c}$ increases significantly with decreasing $f$ ($j_{c}(B=0, T=0) \simeq 0.07$ A/cm$^{2}$ for $f$ = 0.67 and $j_{c}(B=0, T=0) \simeq 0.8$ A/cm$^{2}$ for $f$ = 0.50). The temperature dependence of $j_{c}$ shows strong positive curvature over a wide temperature range for both samples. The phase diagrams of inter- and intragranular superconductivity for different samples indicate that the granular structure might play the key role for an understanding of the origin of superconductivity in the platinum compacts.

\end{abstract}


\section{Introduction} \label{intro}

During the past decades, investigations of superconductors with a non-bulk structure revealed remarkable differences in comparison with the corresponding bulk materials. Lattice disorder in thin films \cite{buckel} as well as a large surface to volume ratio in fine particles made e.g. from aluminum or indium \cite{abeles} caused considerably enhanced critical temperatures arising from particular properties of the phonon spectrum, respectively. In the case of ultrasmall particles, however, electronic quantum size effects may lead to the suppression of superconductivity \cite{peren}. Aspects of granularity play also an important role in the high-$T_{c}$ superconductors since most of those materials have a granular structure, too \cite{deu1}. One particular feature of granular superconductors is the possible occurrence of inter- and intragranular superconductivity in systems consisting of coupled superconducting grains \cite{deu2,deu3,clem,goldfarb}. Intragranular superconductivity denotes the superconducting state of single grains only. Intergranularity means that the supercurrents are not restricted to single or clustered grains, but can flow across grain boundaries and finally even through the whole sample. Due to intergranular coupling effects, a granular superconductor which consists of grains of a type I superconducting material (like, e.g., Al) can effectively behave like a strong type II material \cite{deu1}.
     
Superconductivity of compacted platinum powders was found by measurements of resistivity, ac susceptibility and magnetization (Meissner effect) \cite{ptprl}. In contrast, in bulk platinum no indication for superconductivity has been observed to date down to temperatures of a few $\mu$K \cite{wendi}. Our results on granular platinum reported in Ref. \cite{ptprl} already suggest that this transition concerns the intergranular superconductivity which is responsible for the observation that the electrical resistivity of granular platinum drops to zero in the superconducting phase. In addition, we observe that the magnetic ac susceptibility decreases to a value of about -1; this value of $\chi$ was found to be almost independent of the packing fraction $f$ of the samples ($0.50 \lesssim f \lesssim 0.80$). The ac susceptibility results thus indicate a screening of  the whole platinum compact (bulk and free volume) from external magnetic fields by an intergranular supercurrent flowing circularly around the surface of the cylindrical samples. Hence, for the quest for the origin of superconductivity in granular platinum, a possible separation of inter- and intragranular superconductivity forms an important aspect, as the intragranular effect, i.e. the occurrence of superconductivity in single platinum grains, is supposed to be the precondition for the intergranular effect and therefore should have the more fundamental character. 

\section{Sample Characterization and Experimental Details} 

The educts of the granular platinum samples are commercially available, high purity platinum powders which are the product of a chemical manufacturing process. Our main investigation concentrates on samples made of\, ''Platinum Powder Grade I'' supplied by Alfa Johnson Matthey GmbH, Zeppelinstr. 7, D-76185 Karlsruhe, Germany, and on samples made of a platinum powder from Goodfellow Metals Ltd., Cambridge Science Park, Cambridge CB4 4DJ, England. The high purity of both powders is confirmed by mass spectroscopy. In addition, the content of paramagnetic impurities is investigated by SQUID magnetization measurements at Kelvin temperatures. Assuming the effective magnetic moment of these impurities to be $5 \mu_{B}$ for both powders, we have detected a concentration of $(4\pm 1)$ ppm in the\, ''Alfa powder'' and  $(3\pm 1)$ ppm in the\, ''Goodfellow powder''. Scanning electron microscopy (SEM) studies of both powders revealed the grain size distributions shown in Fig. 1. For the\, ''Alfa powder'', the maximum of the distribution is at a grain diameter of about $1 \mu$m with a distribution width of almost $2 \mu$m (full width at half maximum).The distribution for the\, ''Goodfellow powder'' shows a maximum at $2-3 \mu$m with a width of about $4 \mu$m. Both distributions are asymmetric; the tail of the\, ''Alfa  distribution'' ends at about $4 \mu$m whereas the\, ''Goodfellow distribution'' extends to significantly larger grain sizes (to about $10 \mu$m). The platinum grains in the powders generally are no independent single grains but tend to be clustered together. Moreover, grains with diameters smaller than $0.5 \mu$m which have been found in the\, ''Alfa powder'' exist mostly in groups. 

Mechanical compression of the platinum powders at different pressures $p$ ($1.0 \leq p \leq 4.5$ kbar) results in cylindrical compacts of different packing fraction $f$ ($0.50 \leq f \leq 0.80$), respectively, which is defined as the ratio of the volume of the massive material to the overall sample volume (including voids). Our samples typically have a diameter of 5 mm and a height of 3-4 mm.

Measurements of the magnetic ac susceptibility were performed using the mutual inductance technique (a similar setup is described in ref. \cite{th1}). Up to six samples can be mounted on a silver coldfinger which is attached to the experimental flange of a copper nuclear demagnetization refrigerator \cite{gloos}. Each sample is centered in a separate secondary coil. Two cylindrical superconducting coils which are thermally attached to the mixing chamber of the precooling dilution refrigerator surround the sample holder. They provide the ac primary field as well as static magnetic fields (both in z-direction). The whole setup is surrounded by a superconducting niobium shield. The ambient magnetic field in the environment of the samples is about 70 $\mu$T. This value can be compensated during the cooldown from room temperature by means of a Helmholtz coil system placed around the dewar of the cryostat. 

Phase sensitive measurements of the mutual inductance between the primary coil and the secondary coils which are performed using the AC Inductance and Resistance Bridge LR700, Linear Research Inc., San Diego, USA permit to determine the real part $\chi '$ and the imaginary part $\chi ''$ of the complex magnetic ac susceptibility of the samples. The ac susceptibility is measured at a constant ac frequency of 16 Hz as a function of temperature, static magnetic field and ac excitation field amplitude (given in rms values). The susceptibility values are calibrated versus the change of susceptibility of a AuIn$_{2}$ sample ($T_{c}($AuIn$_{2}) \simeq 208$ mK) at its superconducting transition. All $\chi$ values presented in this work are given in dimensionless SI units. We consider the calibration of the susceptibility values to be accurate within a few percent. The susceptibility data which are generally taken during the warmup of the nuclear stage are computed using the overall sample volume and are corrected for geometric demagnetization \cite{goldfarb,chen1}. The temperature dependent variation of the background signal between 0.1 mK and 30 mK which corresponds to $\chi \simeq 10^{-4}$ can be neglected compared to all changes of susceptibility discussed in this work.

\section{Inter- and Intragranular Effects}
\subsection{Samples made of\, ''Alfa Platinum Powder''}

Figure 2 shows the temperature dependent ac susceptibility of a granular platinum sample (''Alfa platinum'', packing fraction $f$ = 0.67) measured in zero static magnetic field with different ac excitation field amplitudes $b_{AC}$. At temperatures above 1 mK where the electrical resistivity within experimental resolution still has its normal state residual value \cite{ptprl} we already observe a diamagnetic ac susceptibility gradually changing with temperature with a sharp onset at $T \simeq 1.9$ mK (see inset of Fig. 2a). A comparable diamagnetic signal at $T \leq 1.9$ mK was also observed in dc magnetization measurements \cite{ptprl}. Dependent on the ac excitation field, the real part $\chi '$ decreases strongly at $T \lesssim 1$ mK and reaches a value of $\chi ´ \simeq - 0.8$ at the lowest temperatures, whereas the imaginary part $\chi ''$ shows a maximum the position of which depends on the excitation level, too. The smallest ac field amplitude used in our measurements ($b_{AC}$ = 6 nT) turns out to describe approximately the limit of low ac excitation fields. Negative values of $\chi ''$ at low temperatures have no physical meaning and are most likely caused by an experimental artefact (phase shift in the $\chi$ measurement \cite{goldfarb}) which, however, does not affect any of the discussed results. The large decrease of $\chi '$ as well as the dissipative peak in $\chi ''$ are interpreted as the intergranular transition into the superconducting state. The strong dependence of the ac susceptibility on the ac excitation field (nonlinear response) in the regime of intergranular superconductivity is a consequence of the small intergranular critical current density and will be described quantitatively within the framework of the Bean critical state model \cite{bean} in section 5. 

The ac susceptibility between about 1 mK and 1.9 mK  is - in contrast to the behaviour in the intergranular regime - almost independent of the ac excitation field amplitude $b_{AC}$ over a wide interval of $b_{AC}$. Therefore, we conclude that at temperatures above the intergranular transition (but below 1.9 mK) the superconducting critical current density is significantly higher than the intergranular critical current density dominating at lower temperatures. Such a behaviour is expected in granular superconductors consisting of weakly coupled grains, where the intragranular critical current density is much higher than the intergranular one \cite{goldfarb}. Hence, the diamagnetic regime at $1 \lesssim T \lesssim 1.9$ mK can be interpreted in terms of intragranular superconductivity. 

To obtain a sharper separation of inter- and intragranular effects we have investigated another compact consisting of a mixture of a small amount of\, ''Alfa platinum powder'' and ultrafine silver powder (supplied by Tokuriki Honten Co. Ltd., Chiyodaku, Tokyo 101, Japan) with an average grain size of about 100 nm . The platinum powder in this sample has a calculated volume fraction of only about 0.07, and for the silver powder $f = 0.59$ while the residual fraction of 0.34 corresponds to free volume. We should note that these volume fractions are macroscopic parameters referring to the total sample volume and that the platinum powder particles could be packed more densly on a short length scale. However, energy-dispersive X-ray analysis (EDX) with high spatial resolution shows that the platinum clusters are quite homogeneously distributed in the silver powder matrix. 

The temperature dependence of the ac susceptibility of this mixed sample measured at different ac excitation field amplitudes in zero static magnetic field is shown in Fig. 3. The weak temperature dependence of $\chi '$ of the order of $2\cdot 10^{-3}$ at temperatures above 2 mK is caused by paramagnetic impurities in the silver powder \cite{tokpaper}. Compared to the diamagnetic change of $\chi '$ arising from the platinum powder in the superconducting phase this background contribution is negligibly small. Again, we observe intragranular superconductivity at temperatures below about 1.9 mK as in the pure platinum compacts (see Fig. 2). The change of $\chi '$ between $1 \lesssim T \lesssim 1.9$ mK is about $10^{-2}$ and thus one order of magnitude smaller than for the platinum compact with $f$ = 0.67 (compare Figs. 2a and 3a). This indicates that the diamagnetic susceptibility in the intragranular regime scales approximately with the volume fraction of the platinum in the samples. However, in comparison with the pure platinum sample the intergranular superconductivity below about 1 mK is significantly weakened in the mixed Pt/Ag compact: The relative change of $\chi '$ below 1 mK is clearly smaller for the Pt/Ag sample. Neither does $\chi '$ display the characteristic "two step" behaviour nor is there a maximum in $\chi ''$ observable which could be related with the intergranular transition (see Fig. 3b). Obviously, the silver powder acts as normal conducting spacer between the superconducting platinum powder particles and weakens the intergranular supercurrents significantly but does not visibly affect the signal due to intragranular superconductivity. This result supports the description of inter- and intragranular superconductivity in the platinum compacts. The remaining ac excitation field dependence below 1 mK in the Pt/Ag compact (Fig. 3) could be explained by still existing weak intergranular supercurrents between the platinum grains. Proximity-induced superconductivity in the silver powder does presumably not dominate the diamagnetic signal of this sample (for proximity-effects in Ag/Nb systems we refer to \cite{mota}).

\subsection{Samples made of\, ''Goodfellow Platinum Powder''}

A study of the temperature dependent ac susceptibility of a granular platinum compact made of\, ''Goodfellow platinum powder'' ($f$ = 0.52) is displayed in Fig. 4. For this compact we detect the maximum in $\chi ''$ (for $b_{AC}=6$ nT) at  $T \simeq 1.85$ mK and also a sharp onset of diamagnetism at $T \simeq 1.90$ mK (see Fig. 4 and inset). However, as the width of the superconducting transition is significantly smaller for the\, ''Goodfellow samples'' it is hardly possible to separate inter- and intragranular superconductivity for these samples. The sharp transition of $\chi$ in the\, ''Goodfellow samples'' might imply a stronger character of the intergranular superconductivity than in the\, ''Alfa samples'' while the onset of diamagnetism occurring in zero magnetic field at $T \simeq 1.90$ mK might - in analogy to the results on the\, ''Alfa samples'' - also correspond to the onset of intragranular superconductivity. Possibly, the intra- and intergranular transitions in the\, ''Goodfellow samples'' almost coincide due to a relatively strong intergranular coupling of the platinum grains.

It is important to note that in the superconducting regime the value of $\chi '$ at low temparatures is almost $20\%$ larger for the\, ''Goodfellow samples'' than for the\, ''Alfa samples'' (compare Figs. 2 and 4).

\section{Phase Diagrams}  

The transition into the superconducting state of the granular platinum compacts can be suppressed by a static magnetic field. Since the magnitude of the critical magnetic field is comparable to the earth's magnetic field, we have to compensate the residual magnetic field at the sample location by applying a static magnetic field. Fig. 5 shows an example of the magnetic ac susceptibility of a granular platinum sample (''Goodfellow platinum'', $f$ = 0.52) as a function of the applied magnetic field $B_{DC}$  measured at different temperatures. For ''high field'' values the sample is in the normal state whereas at lower fields the (intergranular) superconducting transition occurs which is marked by the drop of $\chi '$ to a value close to -1 and by the maximum in $\chi ''$. The symmetry axis of these $\chi(B_{DC})$ curves at an applied field of about $15 \mu$T corresponds to zero (net) static magnetic field. Since this background field is constant during the whole\, ''run'' of the cryostat the critical magnetic field for the intergranular transition $B_{c}^{inter}$ can be determined from half the distance between two $\chi ''$ peaks (at constant temperature).
 
The intergranular critical magnetic fields as a function of temperature are summarized in Fig. 6a for all samples investigated so far. Measurements of the intergranular critical temperature at constant static magnetic fields (as shown in Figs. 2 and 4) are also plotted in this diagram. We can describe the temperature dependence of the intergranular critical magnetic fields by the equation $B_{c}^{inter}(T) = B_{c0}^{inter}\cdot ( ( 1 - (T/T_{c}^{inter})^{2} )$ for all samples. In general, we find that the intergranular superconducting parameters strongly depend on the packing fraction of the platinum compacts: For the\, ''Alfa platinum'' samples we obtain $6.6  \leq B_{c0}^{inter} \leq 67 \mu$T  and $0.62 \leq T_{c}^{inter} \leq 1.38$ mK for $0.80 \geq f \geq 0.50$, respectively (see also \cite{ptprl});  for the\, ''Goodfellow platinum'' samples: $41 \leq B_{c0}^{inter} \leq 59 \mu$T and $1.55 \leq T_{c}^{inter} \leq 1.85$ mK for $0.66 \geq f \geq 0.52$, respectively (see Tab. 1). Obviously, the intergranular superconducting parameters are also significantly different for samples with same $f$ but made from different platinum powders.

In Fig. 6b we plot the onset of diamagnetic behaviour which we identify with the onset of intragranular superconductivity for all platinum compacts as a function of the static magnetic field. In zero static magnetic field all samples show the onset of intragranular superconductivity at $T \simeq 1.9$ mK. However, one should keep in mind that in the\, ''Goodfellow samples'', inter- and intragranular effects almost coincide in zero magnetic field. Surprisingly, we find that the onset of intragranular superconductivity in a magnetic field depends also on the packing fraction of the samples: In the platinum compacts with the higher packing fractions which show relatively small intergranular critical fields the onset of intragranular superconductivity can also be suppressed more effectively by a static magnetic field.

\section{Intergranular Critical Current Densities}

One pecularity of the ac susceptibility method is the contact free determination of the intergranular critical current density $j_{c}$ which provides important information on the coupling strength among the grains as well as on the supercurrent limiting mechanisms \cite{clem2,darm}. The determination of $j_{c}$ from the ac susceptibility is possible within a critical state model which makes certain assumptions about the flux profile and thus the supercurrent distribution in the superconductor \cite{angurel}.

 We have analysed our ac susceptibility data of superconducting granular platinum within the framework of the Bean critical state model \cite{bean}. This model is applicable at sufficiently low ac excitation fields $b_{AC}$ as it is based on the assumption that $j_{c}(T)$ is independent of the magnetic field \cite{babic}. With $b_{p} = j_{c}\,\mu_{0}\,R$, the following relations hold at $b_{AC} < b_{p}$ for infinitely long cylindrical samples with radius $R$ and their axis parallel to $b_{AC}$ \cite{angurel,babic}:
\begin{equation}
\chi '(b_{AC})  = -1 + \mu_{eff} \cdot \left[\left(\frac{b_{AC}}{b_{p}}\right) - \frac{5}{16} \left(\frac{b_{AC}}{b_{p}}\right)^{2}\right]
\end{equation}
\begin{equation}   
\chi ''(b_{AC}) = \frac{\mu_{eff}}{3\pi} \cdot \left[4 \left(\frac{b_{AC}}{b_{p}}\right) - 2 \left(\frac{b_{AC}}{b_{p}}\right)^{2}\right]         .
\end{equation}

Therefore, $j_{c}$ can be extracted from the linear slope of $\chi '(b_{AC})$ or $\chi ''(b_{AC})$ for small values of $b_{AC}$. The parameter $\mu_{eff} = 1 + \chi_{intra}$ denotes the effective intragranular permeability which is related to the intragranular susceptibility $\chi_{intra}$ arising from the superconducting grains alone without intergranular coupling. From the peak of $\chi ''$ either observed for constant $b_{AC}$ at a temperature $T_{max}$ or at a constant temperature $T$ for a sufficiently strong excitation field $b_{AC} = b_{p}(T)$, $j_{c}$ as a function of temperature can either be obtained using 
\begin{equation}
j_{c}(T_{max}) = b_{AC} / \mu_{0}\,R  
\end{equation} 
or 
\begin{equation}
j_{c}(T) = b_{p}(T) / \mu_{0}\,R  .
\end{equation}
The fact that our samples are no infinitely long cylinders (see section 2) is taken into account by correcting $b_{AC}$ for geometric demagnetization \cite{goldfarb,chen1}.
 
 Assuming that the intergranular critical current density increases with decreasing temperature, one can expect a shift of the maximum in $\chi ''$ to lower temperatures with increasing excitation field $b_{AC}$ (see Eq. 3). This behaviour is shown in Fig. 2 for a granular platinum sample (''Alfa platinum''; $f$ = 0.67). For this sample, we performed also measurements of the ac susceptibility as a function of the ac excitation field $b_{AC}$ at different constant temperatures (Fig. 7). We observe that $\chi '$ and $\chi ''$ both depend linearly on the driving field strength $b_{AC}$ at small $b_{AC}$ as predicted by the Bean model (see Eqs. (1) and (2)). The correction of $b_{AC}$ due to the demagnetization field does not affect significantly these $\chi(b_{AC})$ curves. At higher ac excitation fields (at $b_{AC} = b_{p}$) the maximum in $\chi ''$ and thus the intergranular superconducting transition can be reached. With increasing temperature, the linear slopes of $\chi '$ and $\chi ''$ versus $b_{AC}$ become steeper and the maxima in $\chi ''$ are shifted to lower $b_{p}$, which is again in qualitative agreement with the predictions of the Bean model. For $b_{AC} \gg b_{p}$,  $\chi '$ tends to saturate at finite negative values $\chi_{intra}$, characteristic of the remaining intragranular superconductivity. In this regime, the ac driving field dependence of $\chi$ is significantly smaller than in the intergranular regime presumably because of the higher intragranular critical current density. 

In Fig. 8 we display the $\chi ´´$ data of Figs. 1 and Fig. 7 in a Coles-Coles diagram ($\chi ''$ vs. $\chi '$). The ac excitation field dependent susceptibility data taken at constant temperatures show a linear behaviour of $\chi ''$ versus $\chi '$ at $\chi ' < - 0.4$ (equivalent to $b_{AC} < b_{p}$) with a linear slope which is in good agreement with the theoretical value of $4/3\pi$. The $\chi$ data of Fig. 2 which were measured at constant driving field $b_{AC} = 84$ nT as a function of temperature fall on the expected curve, too (see Fig. 8). However, the susceptibility values taken at $b_{AC} = 18$ nT and $b_{AC} = 6$ nT strongly deviate from this curve. The origin of this deviation is not clear to us. Hence, apart from these two sets of data, the Bean model provides a satisfying description of the excitation field dependent ac susceptibility of the granular platinum sample in the regime of intergranular superconductivity. 

Within the framework of the Bean critical state model, we can thus extract the intergranular critical current density $j_{c}$ from our ac susceptibility measurements. From the $\chi(b_{AC})$ data taken at constant temperatures (see Fig. 7), one can either take the positions of the maxima in $\chi ''$ and apply equation (4) or analyse the shape of $\chi(b_{AC})$ for $b_{AC} < b_{p}$ by taking into account Eqs. (1) and (2).
Both kinds of analysis reveal within resolution the same $j_{c}$ values indicating that a possible magnetic field dependence of $j_{c}$ is presumably not relevant at the comparably low ac field amplitudes used in our experiment \cite{babic}. From Eqs. (1) and (2), however, it is difficult to determine $\mu_{eff}$ (or $\chi_{intra}$) accurately as there is - especially for the measurements performed at low temperatures - hardly any curvature in $\chi(b_{AC})$ for $b_{AC} < b_{p}$. We estimate $\chi_{intra}$ to - 0.14 from the data taken at $T = 0.56$ mK  (Fig. 7) and assume this value also for all other temperatures, since we can consistently describe the $\chi(b_{AC})$ curves with Eqs. (1) and (2) using this value for $\chi_{intra}$. The ac susceptibility data of the investigated Pt/Ag compact (Fig. 3) indicate a saturation of $\chi_{intra}$ at low temperatures, too, which supports the assumption that $\chi_{intra}$ is approximately constant at low temperatures.

We have also performed similar ac susceptibility studies for another\, ''Alfa platinum'' sample with a smaller packing fraction ($f$ = 0.50). All the features of the ac susceptibility explained above were qualitatively observed for this compact, too. In particular, the Bean critical state model is also applicable to the $\chi$ data. In contrast to the compact with $f = 0.67$, we could not observe the maxima in $\chi ''$ at temperatures below 1 mK because our maximum ac excitation field $b_{AC}$ is limited to about $2 \mu$T for technical reasons. For this platinum compact, the values of $j_{c}$ for temperatures below 1 mK  have been calculated from the linear slopes of $\chi(b_{AC})$.

In Fig. 9 the resulting temperature dependent critical current densities at zero static magnetic field are shown for the\, ''Alfa platinum'' samples with packing fractions $f = 0.67$ and $f = 0.50$. Both curves show an increase of $j_{c}$ down to the lowest accessible temperatures. Surprisingly, the absolute values of $j_{c}$ differ by one order of magnitude for the two compacts: We estimate $j_{c}(T=0) \simeq 0.07$ A/cm$^{2}$ for $f = 0.67$ and $j_{c}(T=0) \simeq 0.8$ A/cm$^{2}$ for $f = 0.50$. The temperature dependence of $j_{c}$ is very similar for both samples and shows a strongly positive curvature at $T \gtrsim 0.5\cdot T_{c}^{inter}$. The right axis of Fig. 9 displays the corresponding values of $b_{p} = j_{c}\,\mu_{0}\,R$ which serve as an illustration of how the intergranular superconductivity can be suppressed by the ac driving field: The regions below the $b_{p}$ curves mark the regime of intergranular superconductivity.
 
For both\, ''Alfa platinum'' samples the three data points close to $T_{c}^{inter}$ were obtained from temperature dependent ac susceptibility measurements at constant driving fields (Fig. 2) by analysing the $\chi ''$ maxima and applying Eq. (3). As mentioned above the validity of the Bean model could not be proven for the two $j_{c}$ points close to $T_{c}^{inter}$, respectively. However, those $j_{c}$ values do not significantly affect the shape of the whole $j_{c}$ curves. Furthermore, our choice of the parameter $\mu_{eff}$  for the analysis of the slopes of $\chi(b_{AC})$ does also not significantly affect our results, neither the shape of  $j_{c}(T)$ nor the ratio between the absolute values of $j_{c}$ for the two\, ''Alfa platinum'' samples. We should also mention that the electric transport measurement close to $T_{c}^{inter}$ reported in \cite{ptprl} confirms the order of magnitude of the values of $j_{c}$ presented here which are determined by ac susceptibility measurements. 

Studies of the excitation dependent ac susceptibility of the\, ''Goodfellow platinum'' samples revealed quite different results compared to those on the\, ''Alfa platinum'' samples. Although the superconducting parameters $T_{c}^{inter}$ and $B_{c}^{inter}$ of the\, ''Goodfellow samples'' are comparable to those of the\, ''Alfa samples'' (see Fig. 6 and Tab. 1), their intergranular critical current densities are significantly higher. Unfortunately, due to the experimental limitation in $b_{AC}$ mentioned above, the absolute values of $j_{c}$ for the\, ''Goodfellow samples'' have not been experimentally accessible. From the very weak ac excitation field dependence of $\chi$ at lower temperatures (Fig. 4) we estimate $j_{c}$ - even for the\, ''Goodfellow sample'' with $f = 0.66$ - to be of the order of 10 A/cm$^{2}$ or larger at low temperatures. This result supports qualitatively our picture that in the\, ''Goodfellow platinum'' compacts the platinum grains might be coupled much stronger than those in the\, ''Alfa platinum'' samples.

\section{Discussion}

The characteristic ac excitation field dependence of the ac susceptibility of granular platinum reflects the regimes of inter- and intragranular superconductivity which can be distinguished by significantly different susceptibility values as well as critical current densities. However, an interpretation of our results in terms of a\, ''microscopic'' description of the inter- and especially the intragranular superconductivity in granular platinum is not trivial. Although the method of ac susceptibility measurements enables to distinguish between inter- and intragranular superconductivity, important information about local superconducting properties remains obscured. The total $\chi$ signal of a macroscopic sample in its intragranular superconducting state reflects the sum over all superconducting grains in the sample so that the superconducting properties of one single grain which could, e.g., depend on its size can not be extracted. In particular, the temperature dependent ac susceptibility in the intragranular regime (see Fig. 2 and 3) could arise from a distribution of critical temperatures among the platinum grains. An extreme interpretation of the intra- and intergranular regimes of superconductivity in the platinum compacts could be that the\, ''intragranular'' superconductivity at temperatures close to 1.9 mK arises, e.g., from the fraction of platinum grains with submicron size and that the main amount of platinum grains with a larger grain size which superconduct at lower temperatures are responsible for the intergranular superconductivity. However, taking into account the significant differences between the rather broad grain size distributions of the investigated granular platinum samples (see Fig. 1), it is astonishing that the sharp onset of intragranular superconductivity at $T \simeq 1.9$ mK appears to be a common feature of all our samples and apparently marks the\, ''high temperature'' limit of intergranular superconductivity. 

Another complication of susceptibility measurements on small superconducting grains occurs when the penetration depth of the magnetic field is comparable to the grain size. In that case, the measured susceptibility values would be reduced (''magnetic invisibility'') and could show a temperature dependence arising from the temperature dependent penetration depth \cite{goldfarb,chen2}. This situation might be relevant for our granular platinum samples, in  particular, if superconductivity exists only on the surface of the grains and thus the effective field penetration depth for the grains is relatively large. Hence, one possible striking interpretation of our results and in particular of the sharp onset of intragranular superconductivity at $T \simeq 1.9$ mK (see Figs. 2, 3 and 4) might be the following: At $T \simeq 1.9$ mK the intragranular transition into the superconducting state occurs where basically all the platinum grains become superconducting. Taking into account that the penetration depth ($\propto ((1 - (T/T_{c})^{4})^{-1/2}$ ) \cite{tinkham} below $T_{c} \simeq 1.9$ mK decreases from a value much larger than the grain size to a low temperature value of the order of the grain radius \cite{peren,clem,london}, we could explain the observed decrease of our measured ac susceptibility $\chi_{intra}$ with decreasing temperature from zero (at $T_{c} \simeq 1.9$ mK) to its low temperature value of $- 0.14$ (for the ''Alfa platinum'' sample with $f=0.67$). This picture would also be consistent with the absence of a peak in $\chi ''$ at $T \simeq 1.9$ mK \cite{goldfarb}.

Considerations of the intergranular critical current densities in granular platinum should clearly take into account both the significantly different absolute values of $j_{c}$ for different samples and  the characteristic temperature dependence of $j_{c}$ (see Fig. 9). We observe for samples made from the same platinum powder (''Alfa platinum'') that $j_{c}$ as well as $B_{c}^{inter}$ and $T_{c}^{inter}$ increase significantly with decreasing packing fraction of the compacts. It is surprising that higher packing fractions for which improved electrical coupling of the grains in the normal state is expected appear to be detremental for the intergranular coupling in the superconducting state. Systematic studies of the normal state resistivity of granular platinum samples with different packing fractions have to be performed in order to clarify a possible correlation with the intergranular superconductivity. Such studies might also explain why the\, ''Goodfellow samples'' with similar values for $B_{c}^{inter}$ and $T_{c}^{inter}$ show much higher values of $j_{c}$ than the\, ''Alfa samples''. Considering the grain size distributions of both platinum powders (Fig. 1) as well as our results for compacts with different packing fractions, one may also speculate that in the\, ''Goodfellow powder'' the smaller volume fraction of grains with a diameter smaller than $\simeq 2\, \mu$m is the reason for the observed stronger intergranular superconductivity in these samples.

The temperature dependence of $j_{c}$ (see Fig. 9) can be compared with theoretical predictions for particular supercurrent-limiting mechanisms (see e.g. \cite{darm} and references therein). Assuming that the intergranular supercurrents in the platinum compacts are limited by weak links, the shape of $j_{c}(T)$ can not - at least at temperatures above about $0.5\cdot T_{c}^{inter}$  - be described in terms of the Ambegaokar-Baratoff theory \cite{ambega} for superconductor-insulator-superconductor-type Josephson junctions as this theory predicts a\, ''convex'' $j_{c}(T)$ curve (negative curvature) in contrast to the observed\, ''concave'' behaviour. For a description of the concave $j_{c}(T)$ curve of granular platinum, several theoretical models could be applied qualitatively. An intergrain superconductor-normalconductor-superconductor weak-link structure, but also pair breaking scattering at the grain boundaries would both result in a concave $j_{c}$ curve \cite{degennes,widder}. Moreover, the Ginzburg-Landau theory in the dirty limit predicts a concave slope of  $j_{c}(T)$, too \cite{clem,clem2}. All these supercurrent limiting mechanisms describe\, ''depairing mechanisms''; however, the possibility that depinning of magnetic flux lines is limiting the intergranular supercurrents should also be taken into account. Since in the case of granular platinum the effective coherence length might be of the order of the grain size, intergranular Josephson vortices are supposed to dominate the flux dynamics of the system \cite{clem,tinkham}. Interestingly, pinning of Josephson vortices which might be provided by inhomogenities of the intergranular coupling strengths in the sample could in general be correlated with the\, ''granularity'' (e.g. with the packing fraction $f$). In the case of granular high-$T_{c}$ materials, however, the relevance of pinning of Josephson vortices for the intergranular critical current density appears to be controversial \cite{chen3}. Preliminary measurements of the frequency dependent ac susceptibility indicate flux dynamics effects in the superconducting platinum compacts.

The discussion of inter- and intragranular effects in superconducting granular platinum is closely related with the still unsolved question concerning the origin of superconductivity in this system. The phase diagrams for inter- and intragranular superconductivity in the compacted platinum powders indicate a suppression of superconductivity for more\, ''bulk-like'' samples (samples with higher packing fractions). Hence, the fact that superconductivity has not yet been observed in bulk platinum along with the phase diagrams and the intergranular critical current densities for superconducting granular platinum suggest that the granular structure of the platinum compacts might play a dominant role for the occurrence of superconductivity. Possibly, ''lattice softening'' due to the large surface to volume ratio might cause an enhanced transition temperature \cite{abeles}.
On the other hand, the strongly exchange enhanced paramagnetism which is supposed to prevent the superconducting pairing in bulk platinum \cite{gladstone} as well as impurity magnetism should be taken into account, too. The extremely weak impurity magnetism observed at mK temperatures \cite{ptprl} appears to be at least an indication for a relation between the granularity of the samples and their magnetic properties which could also play an important role for the superconductivity.  

A microscopic description of inter- and intragranular superconductivity in this system can hardly be given at the moment. In particular, it is not yet clear what the superconducting\, ''grains'' actually are. It has to be clarified whether the intragranular superconductivity corresponds to superconductivity of single platinum grains, or to clusters of grains and whether these clusters are the prerequisite for the occurrence of superconductivity. Moreover, it would be important to know whether the superconductivity of these grains is bulk-like or whether it exists only on the surface of the grains. For the understanding of the intergranular superconductivity in granular platinum, the understanding of the intragranular effect should evidently be a prerequisite as the latter effect is supposed to be the precondition for the former. Our phase diagrams (Fig. 6) indicate a common onset of intragranular superconductivity at $T \simeq 1.9$ mK in zero magnetic field. At higher magnetic fields, however, we observe a suppression of inter- and intragranular superconductivity with increasing packing fraction, suggesting a correlation between intra- and intergranular effects but also an effect of the\, ''granularity'' on both types of superconductivity. Furthermore, the local magnetic field close to the platinum grains might differ significantly from the externally applied magnetic field \cite{askew} and might thus be also an important factor for the appearance of intra- as well as intergranular superconductivity.

To summarize, our investigations of the excitation field dependent ac susceptibility along with the measurements of the resistivity and magnetization \cite{ptprl} of superconducting compacted platinum powders reveal a possible separation of inter- and intragranular effects in this system. However, the still unclear origin of superconductivity in granular platinum does not yet allow a microscopic description of the intra- and intergranular effects. The phase diagrams for inter- and intragranular superconductivity as well as the results for intergranular critical current densities indicate that the granular structure of the platinum compacts might play a dominant role for the occurrence of superconductivity. Hence, further detailed studies of the impact of the topology of various platinum samples especially on their superconducting and magnetic properties should provide a key for a deeper understanding of the superconductivity in granular platinum.

\subsection*{Acknowledgements}
We are grateful for the technical contributions of I. Usherov-Marshak, A. Dertinger, C. Drummer (BIMF Bayreuth) and F. Siegelin (IMA Bayreuth). This work was supported through DFG grants Ko1713/1-3, Ko1713/6-1 and He2282/2-1,2 as well through the EU-TMR Large Scale Facility project (contract no. ERBFMGECT950072).

\begin{figure}[h]
\centerline{\includegraphics[width=0.6\textwidth]{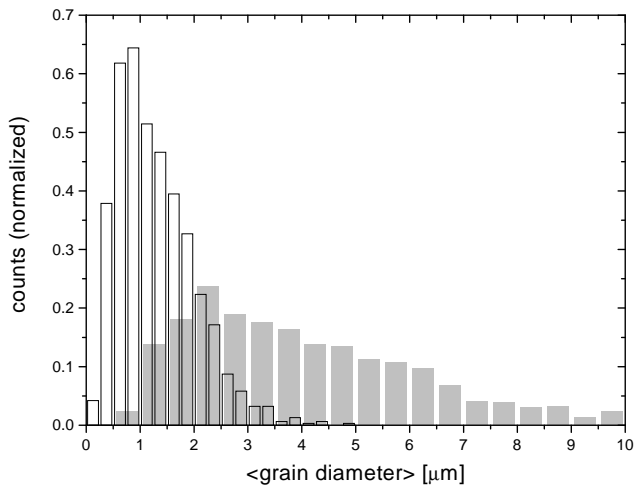}}
\caption{Grain size distributions of the ''Alfa platinum powder'' (open columns) and of the\, ''Goodfellow platinum powder'' (grey columns) obtained from SEM studies.}
\end{figure}
00

\begin{figure}
\centerline{\includegraphics[width=0.5\textwidth]{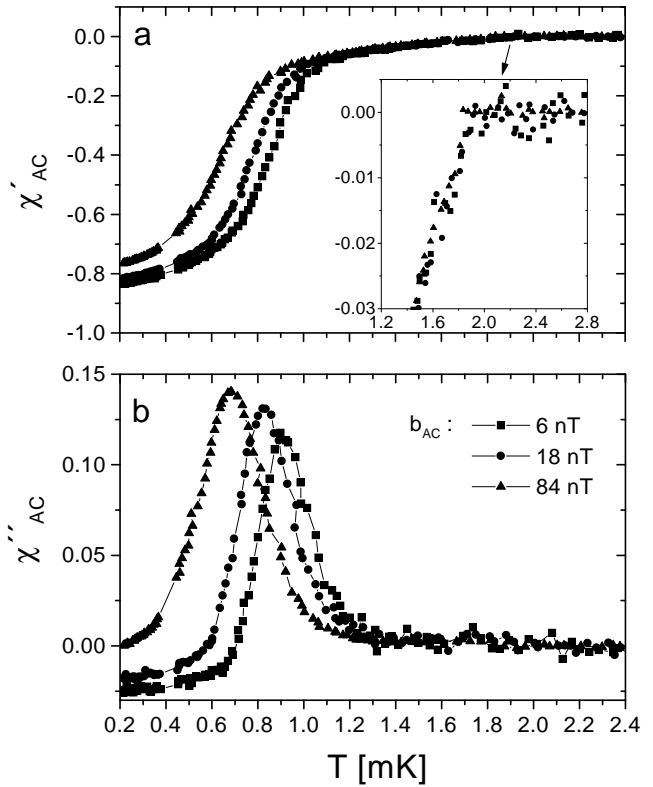}}
\caption{Temperature dependence of the magnetic ac susceptibility ((a) $\chi '$ and (b) $\chi ''$) of granular platinum (''Alfa platinum''; packing fraction $f = 0.67$) in zero static magnetic field measured with different ac excitation field amplitudes $b_{AC}$. The strongly negative and excitation field dependent ac susceptibility  below 1 mK corresponds to intergranular superconductivity whereas the ac driving field independent regime at $1 \lesssim T \lesssim 1.9$ mK is interpreted in terms of intragranular superconductivity. Note the sharp onset of the intragranular regime at $T \simeq 1.9$ mK (see inset in figure a).}
\end{figure}

\begin{figure}
\centerline{\includegraphics[width=0.5\textwidth]{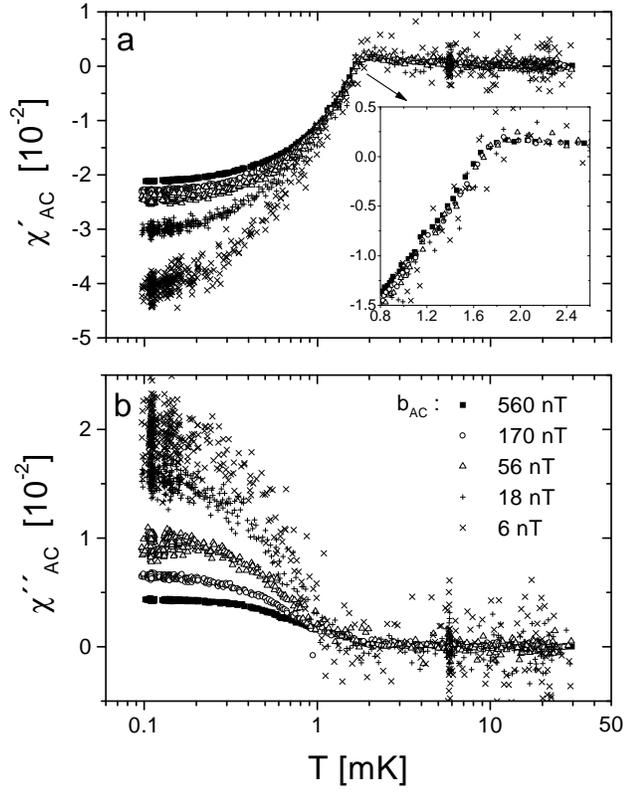}}
\caption{Temperature dependence of the magnetic ac susceptibility of a compact of mixed ''Alfa platinum'' and silver powders (volume fractions: $f_{Pt}$ = 0.07, $f_{Ag}$ = 0.59) in zero static magnetic field taken at different ac excitation field amplitudes $b_{AC}$. The inset in figure (a) shows the almost ac driving field independent intragranular regime with its sharp onset at $T \simeq 1.9$ mK.}
\end{figure}

\begin{figure}
\centerline{\includegraphics[width=0.5\textwidth]{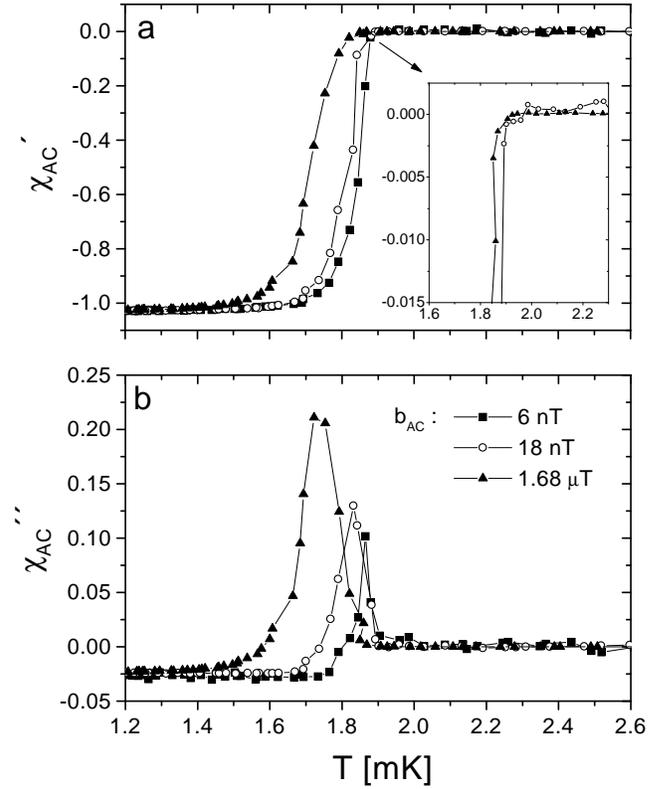}}
\caption{Temperature dependent magnetic ac susceptibility of granular platinum (''Goodfellow platinum''; packing fraction $f$ = 0.52) in zero static magnetic field at different ac excitation field amplitudes $b_{AC}$. The inset of figure (a) shows the onset of diamagnetic behaviour at $T \simeq 1.9$ mK.}
\end{figure} 

\begin{figure}
\centerline{\includegraphics[width=0.5\textwidth]{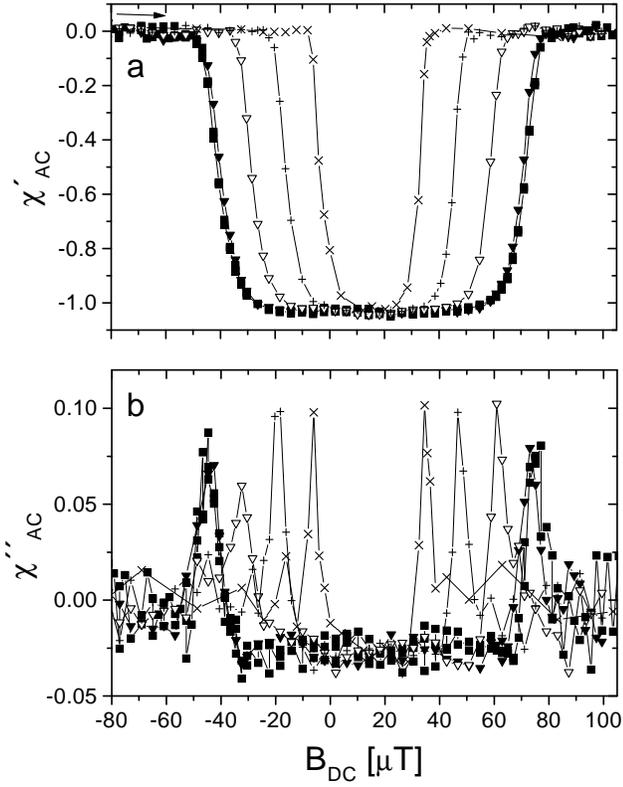}}
\caption{Magnetic ac susceptibility of granular platinum (''Goodfellow platinum''; packing fraction $f$ = 0.52) as a function of the applied static magnetic field $B_{DC}$ at temperatures $T = 0.1$ mK ($\blacksquare$), 0.24 mK ($\blacktriangledown$), 0.85 mK ($\triangledown$), 1.22 mK ($+$) and 1.51 mK ($\times$) measured with an ac excitation field amplitude of $b_{AC}$ = 6 nT.}
\end{figure}

\begin{figure}
\centerline{\includegraphics[width=0.5\textwidth]{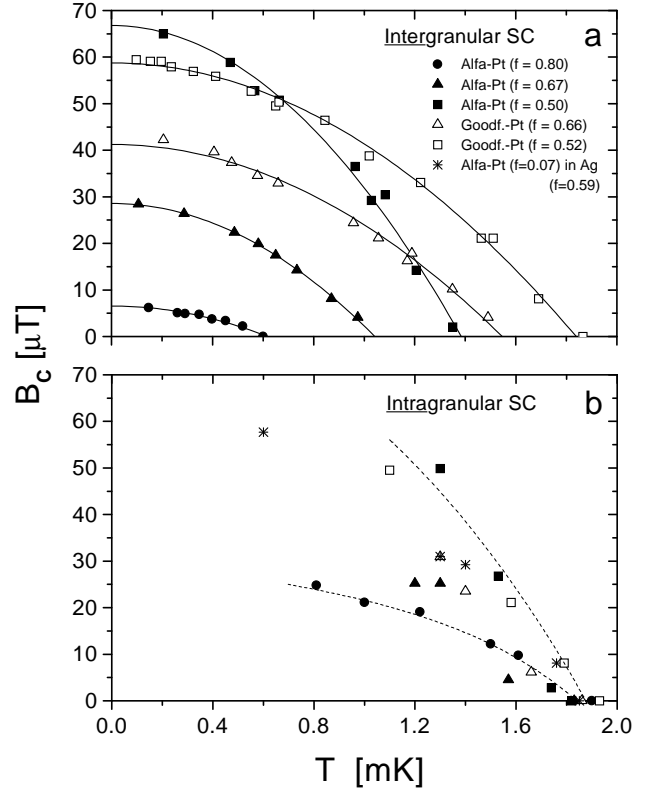}}
\caption{(a) Intergranular critical magnetic fields of various platinum compacts as a function of temperature. The lines are fits to the function $B_{c}(T) = B_{c0}\cdot((1- (T/T_{c})^{2})$.\\ 
(b) Onset temperatures of intragranular superconductivity as a function of the static magnetic field. The dashed lines are guides to the eye.}
\end{figure}
 
\begin{figure}
\centerline{\includegraphics[width=0.5\textwidth]{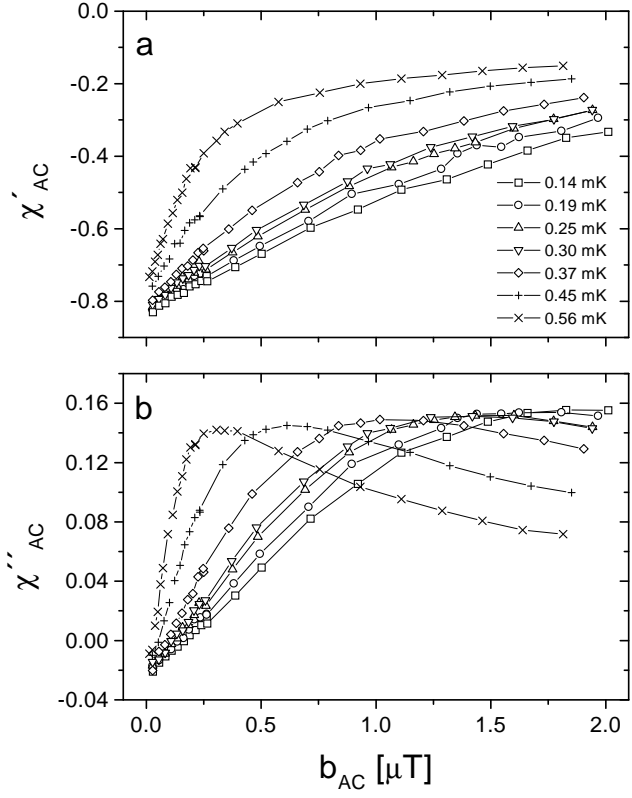}}
\caption{Dependence of the magnetic ac susceptibility of granular platinum (''Alfa platinum''; packing fraction $f$ = 0.67) on the ac excitation field $b_{AC}$ for different constant temperatures. The measurements were taken in zero static magnetic field.}
\end{figure}

\begin{figure}
\centerline{\includegraphics[width=0.55\textwidth]{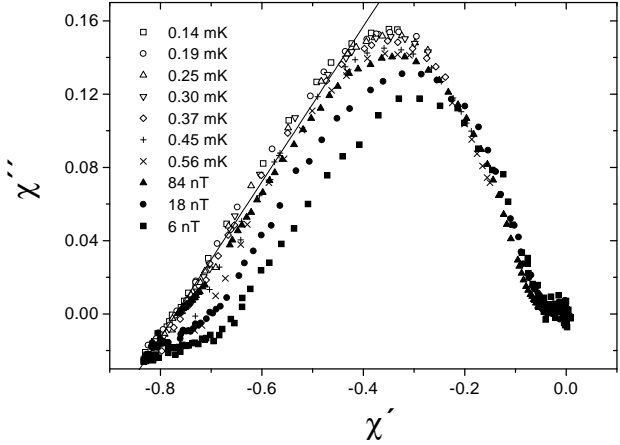}}
\caption{Plot of $\chi ''$ versus $\chi '$ (Coles-Coles diagram) for granular platinum (''Alfa platinum''; packing fraction $f$ = 0.67) in zero static magnetic field. The open symbols refer to the $\chi(b_{AC})$ data from Fig. 7 measured at constant temperatures; the full symbols refer to the $\chi$ data from Fig. 2 measured as a function of temperature at constant $b_{AC}$. The solid line indicates the theoretically expected slope of $4/3\pi$.}
\end{figure}

\begin{figure}
\centerline{\includegraphics[width=0.55\textwidth]{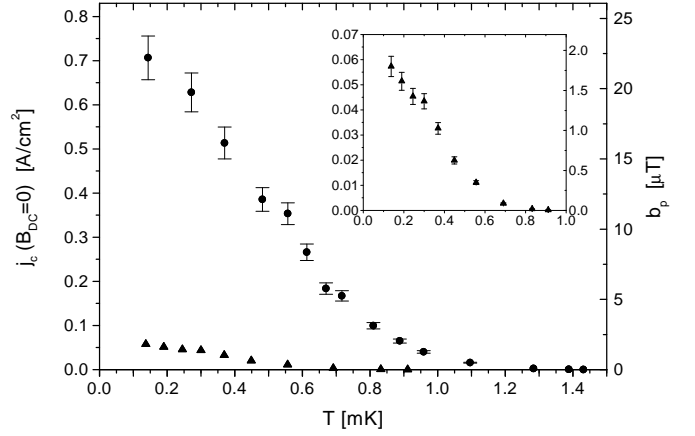}}
\caption{Intergranular critical current densities $j_{c}$ in zero static magnetic field (obtained from ac excitation field dependent susceptibility measurements) for\, ''Alfa platinum'' samples with the packing fractions $f = 0.50$ ($\bullet$) and $f = 0.67$ ($\blacktriangle$) (see also inset). The right axis shows $b_{p} = j_{c}\,\mu_{0}\,R$. Note the significantly different absolute values of $j_{c}$ for the two samples and the strongly positive curvature of the temperature dependence of $j_{c}$ over a wide temperature range for both samples.}
\end{figure}

\begin{table}[h]
\begin{center}
\begin{tabular}{c|c|c|c}
supplier & $f$ &
$T_{c}^{inter}$ [mK] &
$B_{c0}^{inter} \, [\mu$T] \\ \hline

Alfa Johnson Matthey GmbH & $0.80$ & 0.62 & $6.6$ \\ \hline
'' & $0.67$ & 1.04 & $29$  \\ \hline
'' & $0.50$ & 1.38 & $67$  \\ \hline
Goodfellow Metals Ltd. & $0.66$ & 1.55 & $41$ \\ \hline
'' & $0.52$ & 1.85 & $59$  
\end{tabular}\end{center}
\caption{Intergranular superconducting parameters $T_{c}^{inter}$
and $B_{c0}^{inter}$ for various compacted Pt powders with different packing 
fractions $f$.}
\end{table}


\begin{references}

\bibitem{ptprl}R. K\"onig, A. Schindler, and T. Herrmannsd\"orfer, {\it Phys. Rev. Lett.} {\bf 82}, 4528 (1999)

\bibitem{buckel}W. Buckel and R. Hilsch, {\it Z. Physik} {\bf 131}, 420, (1952), {\it Z. Physik} {\bf 138}, 109, (1954) 

\bibitem{abeles}B. Abeles, {\it Appl. Solid State Sci.} {\bf 6}, 64 (1976);
      S. Matsuo, H. Sugiura, and S. Noguchi, {\it J. Low Temp. Phys.} {\bf 15}, 481 (1973)

\bibitem{peren}J.A.A.J. Perenboom, P. Wyder, and F. Meier, {\it Physics Reports} (Review Section of Physics Letters) {\bf 78}, 173 (1981);
D.C.  Ralph, C.T. Black, and M. Tinkham, {\it Phys. Rev. Lett.} {\bf 78}, 4087 (1997)

\bibitem{deu1}G. Deutscher, {\it Mat. Res. Soc. Symp. Proc.} {\bf 195}, 303 (1990)

\bibitem{deu2}G. Deutscher, H. Fenichel, M. Gershenson, E. Gr\"unbaum, and Z. Ovadyahu, {\it J. Low Temp. Phys.} {\bf 10}, 231 (1973)

\bibitem{deu3}G. Deutscher, Y. Imry, and L. Gunther, {\it Phys. Rev. B} {\bf 10}, 4598 (1974)

\bibitem{clem}J.R. Clem, {\it Physica C} {\bf 153}, 50 (1988)

\bibitem{goldfarb}R.B. Goldfarb, M. Lelental, and C.A. Thompson, in {\it Magnetic Susceptibility of Superconductors and Other Spin Systems}, edited by R.A. Hein et al., Plenum Press, New York, (1991)

\bibitem{wendi}W. Wendler, T. Herrmannsd\"orfer, S. Rehmann, and F. Pobell, {\it Europhys. Lett.} {\bf 38}, 619 (1997)

\bibitem{th1}T. Herrmannsd\"orfer, S. Rehmann, and F. Pobell, {\it J. Low Temp. Phys.} {\bf 104}, 67 (1996)

\bibitem{gloos}K. Gloos, P. Smeibidl, C. Kennedy, A. Singsaas, P. Sekowski, R.M. Mueller, and F. Pobell, {\it J. Low Temp. Phys.} {\bf 73}, 101 (1988)

\bibitem{chen1}D.-X. Chen, J.A. Brug, and R.B. Goldfarb, {\it IEEE Trans. Magn.} {\bf 27}, 3601 (1991)

\bibitem{bean}C.P. Bean, {\it Phys. Rev. Lett.} {\bf 8}, 250 (1962); C.P. Bean, {\it Rev. Mod. Phys.} {\bf 36}, 31 (1964)


\bibitem{tokpaper}I. Usherov-Marshak, A. Schindler, and R. K\"onig, {\it Physica B} {\bf 284-288}, 202 (2000)

\bibitem{mota}A.C. Mota, P. Visani, and A. Pollini, {\it J. Low Temp. Phys.} {\bf 76}, 465 (1989);
P. Visani, A.C. Mota, and A. Pollini, {\it Phys. Rev. Lett.} {\bf 65}, 1514 (1990)

\bibitem{clem2}J.R. Clem, B. Bumble, S.I. Raider, W.J. Gallagher, and Y.C. Shih, {\it Phys. Rev. B} {\bf 35}, 6637  (1987)

\bibitem{darm}H. Darhmaoui and J. Jung, {\it Phys. Rev. B} {\bf 53}, 14621 (1996)

\bibitem{angurel}L.A. Angurel, F. Lera, C. Rillo, and R. Navarro, {\it Physica C} {\bf 230}, 361 (1994) 

\bibitem{babic}Z. Marohnic and E. Babic, in {\it Magnetic Susceptibility of Superconductors and Other Spin Systems}, edited by R.A. Hein et al., Plenum Press, New York, (1991)

\bibitem{chen2}D.-X. Chen, A. Sanchez, T. Puig, L.M. Martinez, and J.S. Munoz, {\it Physica C} {\bf 168}, 652 (1990)

\bibitem{london}F. London, {\it Superfluids} , Vol. {\bf 1}, Dover Publications, New York, (1961)   

\bibitem{tinkham}M. Tinkham, {\it Introduction to Superconductivity}, 2nd edition, McGraw-Hill, New York, (1996)

\bibitem{ambega}V. Ambegaokar and A. Baratoff, {\it Phys. Rev. Lett.} {\bf 10}, 486 (1963), and {\bf 11}, 104(E) (1963)

\bibitem{degennes}P. G. DeGennes, {\it Rev. Mod. Phys.} {\bf 36}, 225 (1964)

\bibitem{widder}W. Widder, L. Bauernfeind, H.F. Braun, H. Burkhardt, D. Rainer, M. Bauer, and H. Kinder, {\it Phys. Rev. B} {\bf 55}, 1254 (1997)

\bibitem{chen3}D.X. Chen, J.J. Moreno, A. Hernando, and A. Sanchez, to be published in {\it Physica C} (Proc. M2S-HTSC-VI, Houston, Texas, USA, February 2000
 
\bibitem{gladstone}G. Gladstone, M.A. Jensen, and J.R. Schrieffer, Superconductivity in the transition metals, in {\it Superconductivity}, R.D. Parks (ed.), Vol II, 665 (1969)

\bibitem{askew}T. R. Askew, R. Flippen, K. J. Leary, and M.N. Kunchur, {\it J. Mater. Res.} {\bf 6}, 1135 (1991)
\end{references}
\end{document}